\documentclass[11pt,twoside]{article}

\usepackage{asp2006}
\usepackage{epsf}
\usepackage{psfig}
\usepackage{lscape}
\usepackage{epsfig,subfigure,amsmath,amssymb}

\begin{document}

\title{Homography-based correction of positional errors in MRT survey}
\author{Arvind Nayak$^*$, Soobash Daiboo$^{*,\dagger}$ and N. Udaya Shankar$^*$}
\affil{$*$ Raman Research Institute, Bangalore 560 080, INDIA\\
$^\dagger$ Physics Department, University of Mauritius, MAURITIUS\\
{\tt Email: \{arvind, soobash, uday\}@rri.res.in} }

\begin{abstract}
The Mauritius Radio Telescope (MRT) images show systematics in 
the positional errors of sources when compared to source positions in
the Molonglo Reference Catalogue (MRC). We have applied two-dimensional 
homography to correct positional errors in the image domain and avoid 
re-processing the visibility data. Positions of bright (above 15-$\sigma$) 
sources, common to MRT and MRC catalogues, are used to set up an over-determined 
system to solve for the 2-D homography matrix. After correction, the errors are 
found to be within 10\% of the beamwidth for these bright sources and the 
systematics are eliminated from the images.
\end{abstract}

\vspace{-28pt}
\section{Introduction}
\label{s:introduction}

Images of about a steradian ($18^{\mbox{h}} \leq \alpha \leq 
24^{\mbox{h}}30^{\mbox{m}}\mbox{, }-70^{\circ} \leq \delta \leq 
-10^{\circ}$) of the southern sky, at 151.5~MHz, 
using MRT~\citep{apss:uday02}, were produced by~\cite{ursiga:pandey05}. 
The deconvolved images, with an angular resolution of 
$4'\times4'.6\mbox{ sec (} \delta + 20^{\circ}.14\mbox{)}$ and rms noise of 
about 300~mJy~beam$^{-1}$, and a source catalogue of 2782 sources were 
published by~\cite{thesis:pandey06}. 

\begin{figure}[!t]
\centering
\subfigure[]{
\epsfig{figure=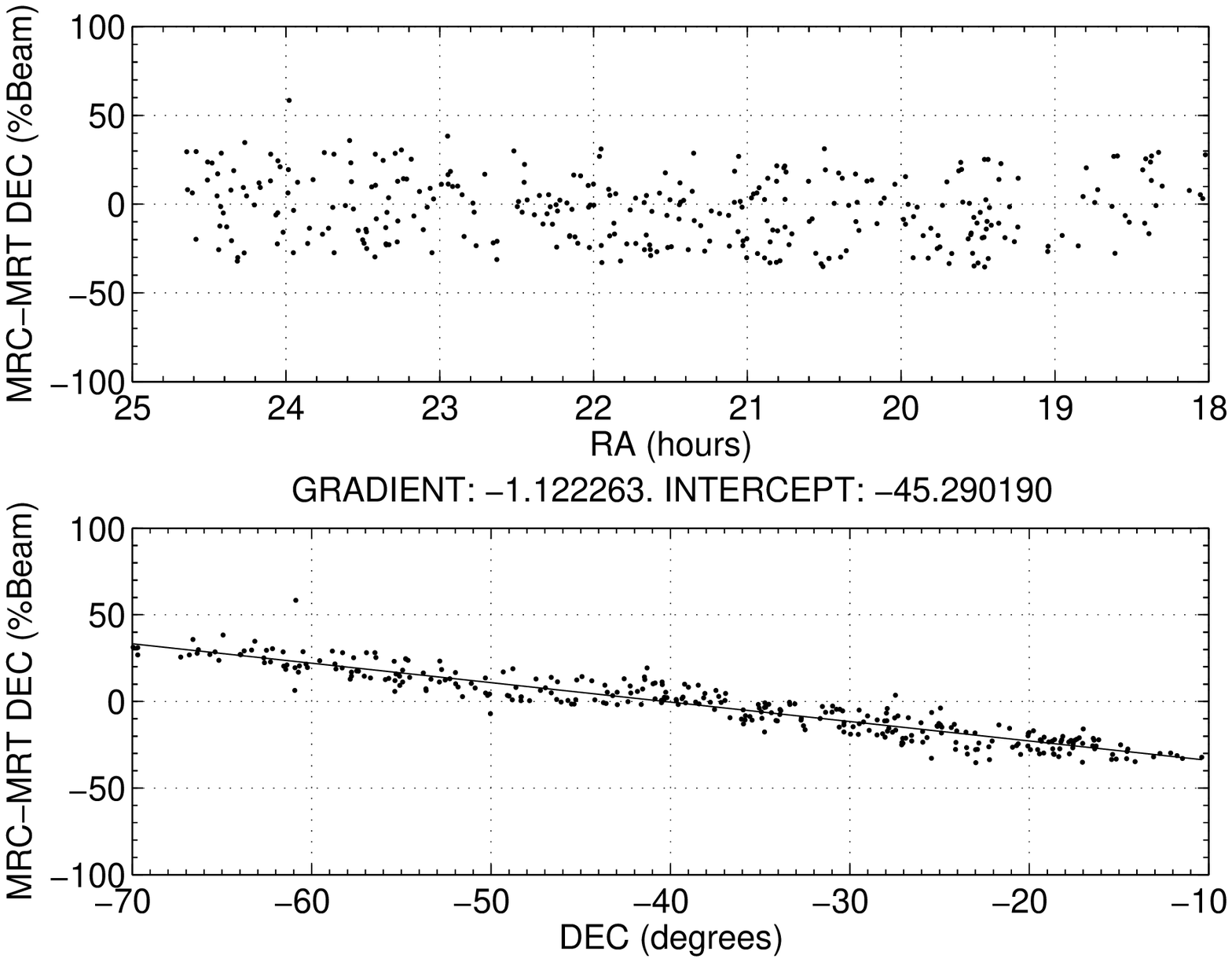,width=0.4\linewidth}}
\subfigure[]{
\epsfig{figure=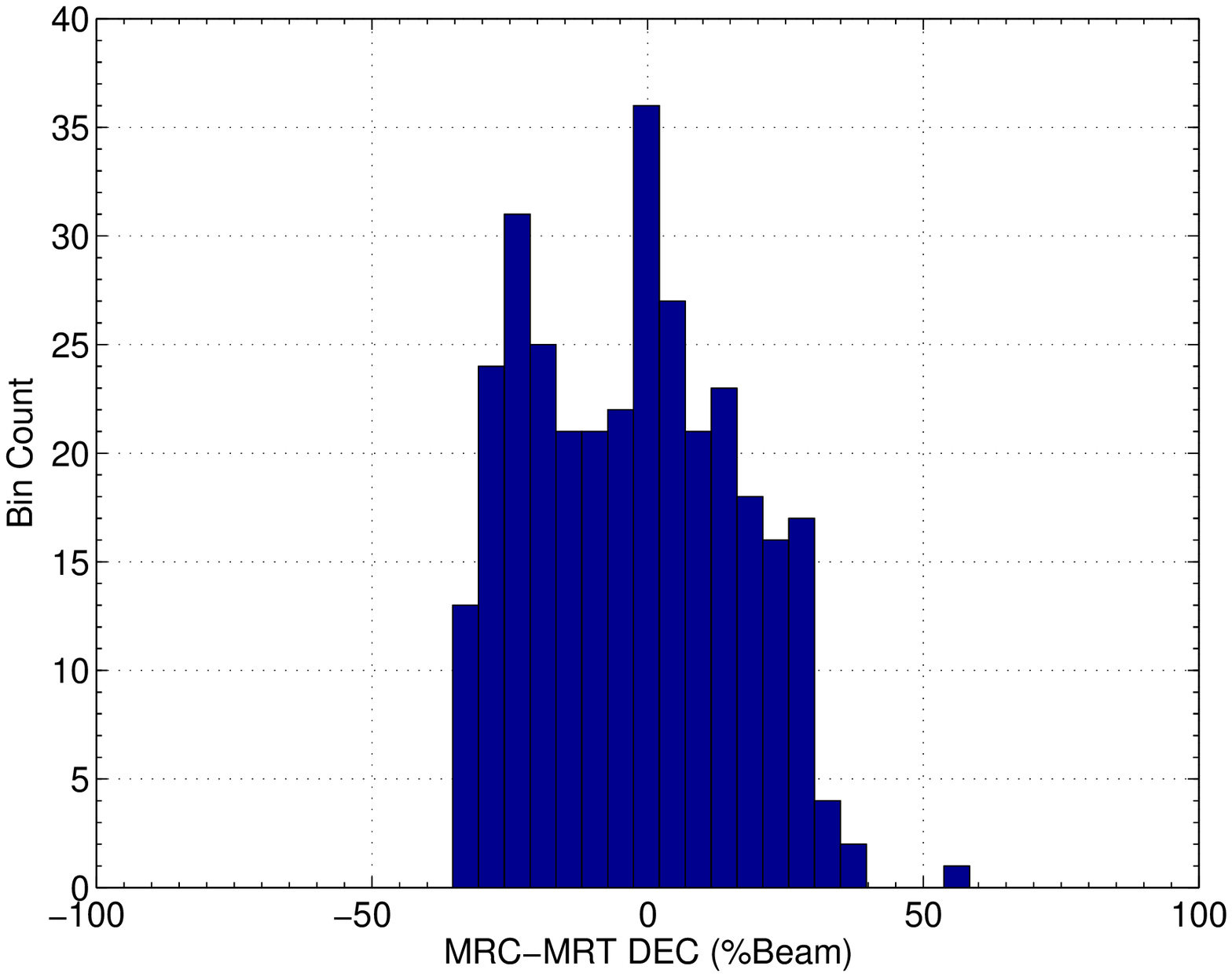,width=0.4\linewidth}}
\caption{ \small 
Positional error analysis. (a) Plot in the first row shows DEC errors 
against RA; no systematics are observed. Second row plot shows DEC 
errors against DEC; errors show a linear gradient as a function 
of DEC. (b) Histogram of positional errors in DEC. \normalsize }
\vspace{-5pt}
\label{f:sourcecomparison}
\end{figure}

Systematics in positional errors were found when the positions of sources
common to MRT and MRC \citep{mnras:large81} catalogues were compared. The 
positional errors in right ascension (RA) show no systematics as a function 
of RA and the errors are within 15\% of the RA beamwidth. The positional 
errors in declination (DEC) also show no systematics as a 
function of RA. However, errors in DEC show a linear gradient as a 
function of DEC (Figure~\ref{f:sourcecomparison}(a)) and reach about 
50\% of the MRT beamwidth in declination. Histogram in 
Figure~\ref{f:sourcecomparison}(b) shows the distribution of errors in DEC. 
Due to space limitations similar plots for RA errors are not shown. 
~\cite{thesis:pandey06} used one-dimensional (1-D) robust least squares fit, 
estimated the errors and corrected them in the source catalog. However, 
the errors remained in the images, which impede usefulness of MRT images for 
multi-wavelength analysis of sources.

This paper describes the application of two-dimensional (2-D) homography, 
a technique ubiquitous in the computer vision and graphics community, to 
correct the errors in the image domain. At MRT, the visibility data is 
processed through several complex stages of data reduction specific to the 
array, especially, arising due to its non-coplanarity~\citep{apss:uday02}. 
It was therefore decided to correct for the errors in the image domain and 
avoid re-processing the visibility data. Homography is used to estimate a 
transformation matrix (which includes rotation, translation and non-isotropic 
scaling) that accounts for positional errors in the linearly gridded 2-D images. 

\vspace{-10pt}
\section{Homography-based correction scheme}
\label{s:homography}

2-D planar homography is a non-singular linear relationship between
points on planes. Given two sets of corresponding image points in 
projective coordinates, ${\bf p}, {\bf p'} \in \mathbb{P}^{2}$,
homography maps each ${\bf p}$ to the corresponding ${\bf p'}$ 
\citep{book:hartley}. The homography sought is a non-singular 
$3\times 3$ matrix $\tt H$ such that:
\begin{equation}
\left[ \begin{array}{c}
x^{'}\\y^{'}\\1 
\end{array}\right]=
\left[ \begin{array}{ccc}
h_{11}&h_{12}&h_{13}\\
h_{21}&h_{22}&h_{23}\\
h_{31}&h_{32}&h_{33}
\end{array}\right]
\left[ \begin{array}{c}
x\\y\\1\end{array}\right].
\label{eq:inhomog}
\end{equation}
Where, $(x,y)$ and $(x',y')$ represent (RA, DEC) of
MRT and MRC sources, respectively. 
$\tt H$ is defined up to a scale factor, hence the problem has 
8 degrees of freedom. Each point provides two equations, so a 
minimum of 4 point correspondences are necessary to fully 
constrain $\tt H$. The equation contributed by each corresponding 
pair can be re-arranged to obtain:
\begin{equation}
\label{twoequations}
\begin{array}{c}
x^{'}(h_{31}x+h_{32}y+h_{33})=h_{11}x+h_{12}y+h_{13}\\
y^{'}(h_{31}x+h_{32}y+h_{33})=h_{21}x+h_{22}y+h_{23}.
\end{array}
\end{equation}
A set of $n$ such equation pairs, contributed by $n$ point correspondences, 
form an over-determined linear system ${\tt A}{\mathbf h}={\mathbf b}$, as 
shown in Eq.~\ref{e:system}.
\begin{equation}
\label{e:system}
\left[ \begin{array}{cccccccc}
x_{1} & y_{1} & 1 & 0 & 0 & 0 & -x_{1}x_{1}' & -x_{1}'y_{1}\\
0 & 0 & 0 & x_{1} & y_{1} & 1 & -x_{1}y_{1}' & -y_{1}y_{1}'\\
\vdots &&&& \vdots&&&\vdots\\
x_{n} & y_{n} & 1 & 0 & 0 & 0 & -x_{n}x_{n}' & -x_{n}'y_{n}\\
0 & 0 & 0 & x_{n} & y_{n} & 1 & -x_{n}y_{n}' & -y_{n}y_{n}'\\
\end{array}\right]
\left[ \begin{array}{ccc}
h_{11}\\
h_{12}\\
h_{13}\\
h_{21}\\
h_{22}\\
h_{23}\\
h_{31}\\
h_{32}
\end{array}\right]=
\left[ \begin{array}{c}
x_{1}'\\y_{1}'\\
\vdots\\
x_{n}'\\y_{n}'\\
\end{array}\right]
\end{equation}

\begin{figure}[!t]
\centering
\epsfig{figure=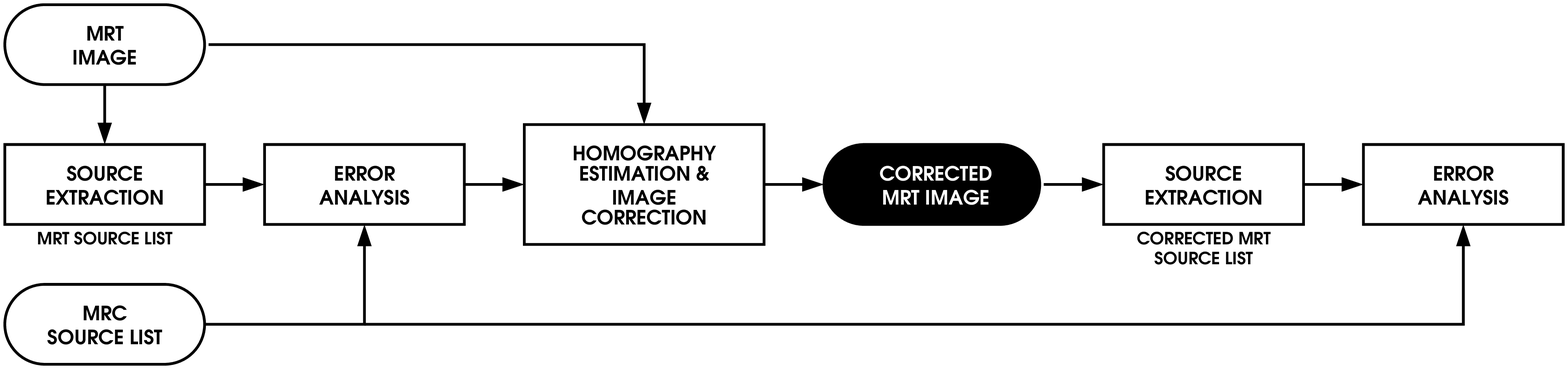,width=\linewidth}
\caption{ \small Block schematic of the correction scheme. Rectangular
boxes represent processes, rounded boxes data or results. \normalsize }
\label{f:scheme}
\vspace{-5pt}
\end{figure}

This system can be solved by least squares based estimators.
To obtain a good estimate of the transformation matrix we 
adopted the normalization scheme proposed by \cite{pami:hartley97}. 
This ensures freedom on arbitrary choices of scale and coordinate 
origin, leading to algebraic minimization in a fixed 
canonical frame.

Figure~\ref{f:scheme} shows the block schematic of the correction 
scheme. We identify bright (above 15-$\sigma$) sources common to the 
MRT and MRC catalogues. There are about 400 common sources in the 
steradian of the sky under consideration. A single $3\times3$ 
homography matrix (refer equation~\ref{eq:inhomog}) was estimated from
this population. Each pixel from 
the images was then projected to a new position using this
matrix. 

\vspace{-10pt}
\section{Results}

\begin{figure}[!t]
\centering
\subfigure[]{
\epsfig{figure=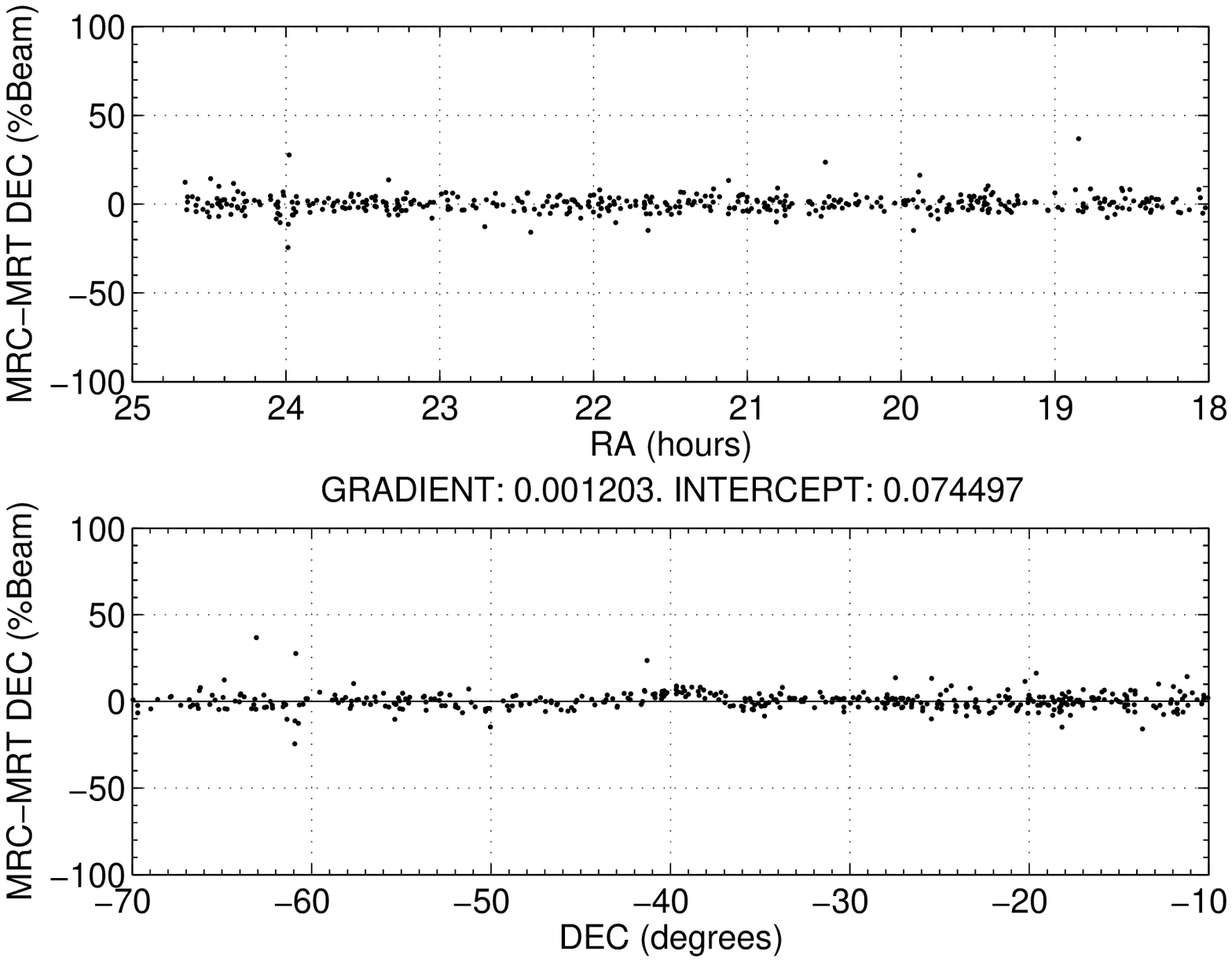,width=0.4\linewidth}}
\subfigure[]{
\epsfig{figure=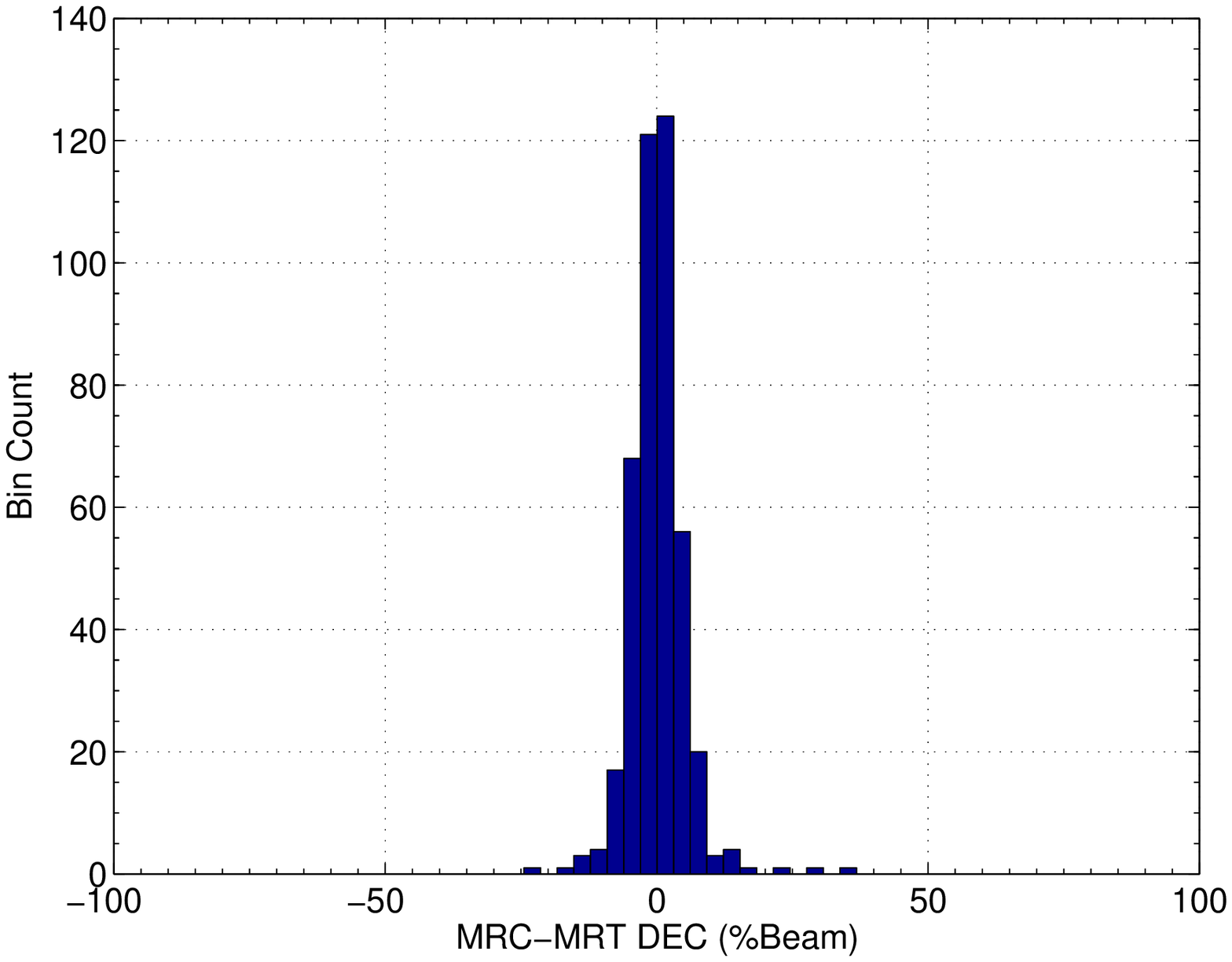,width=0.4\linewidth}}
\caption{\small Positional error analysis after 
homography-based correction. (a) First and
second rows show DEC errors against RA and DEC,
respectively. (b) Histogram of DEC positional 
errors. \normalsize }
\vspace{-5pt}
\label{f:sourcecomparisoncorrect}
\end{figure}

Figure~\ref{f:sourcecomparisoncorrect} shows positional errors in DEC
after homography-based correction. Compare these plots with 
Figure~\ref{f:sourcecomparison}. Clearly, homography has managed to
remove systematics and reduce the errors to well within 10\% of the 
beamwidth for sources above 15-$\sigma$.

\begin{figure}[!t]
\centering
\subfigure[]{
\epsfig{figure=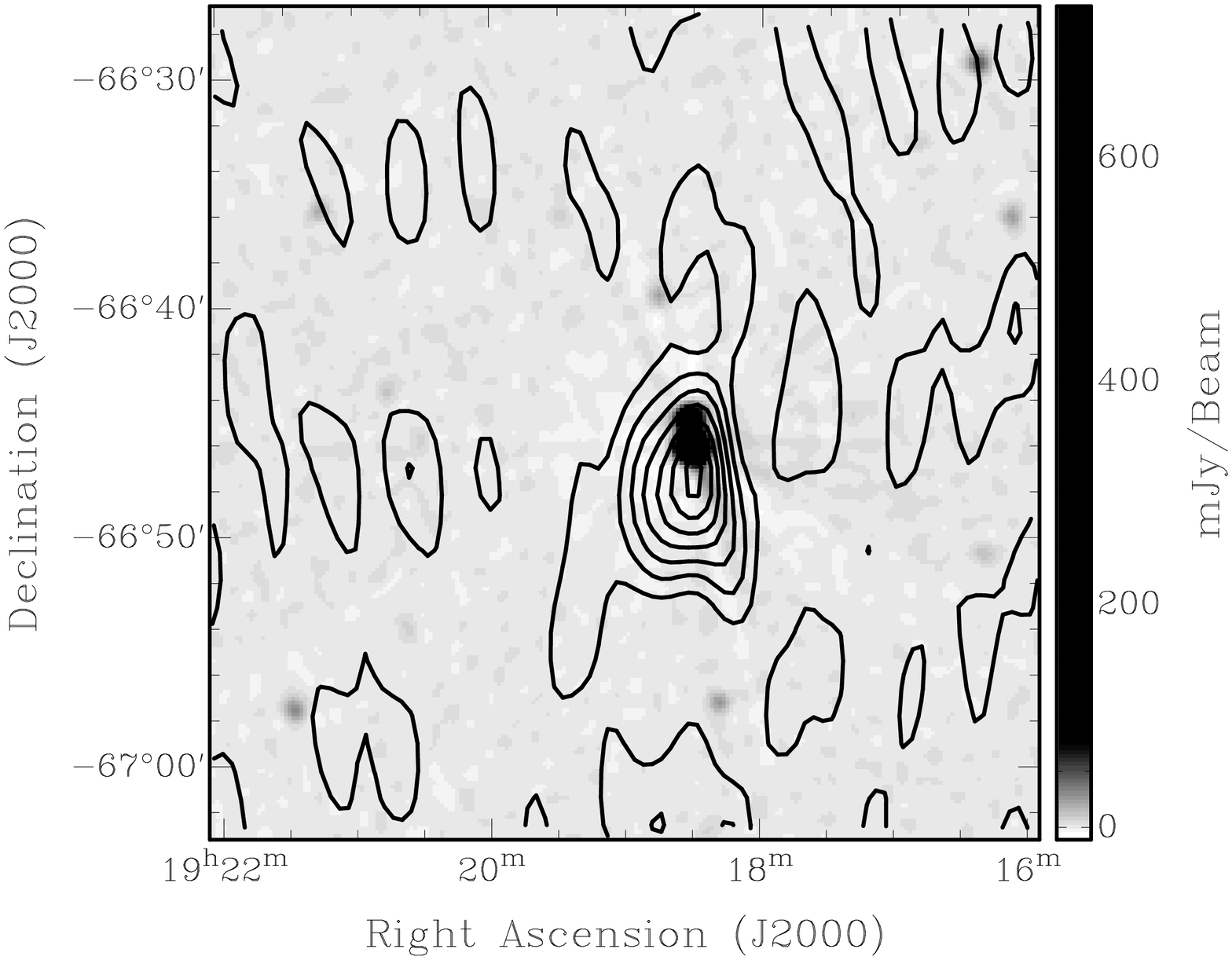,angle=-90,width=0.41\linewidth}}
\subfigure[]{
\epsfig{figure=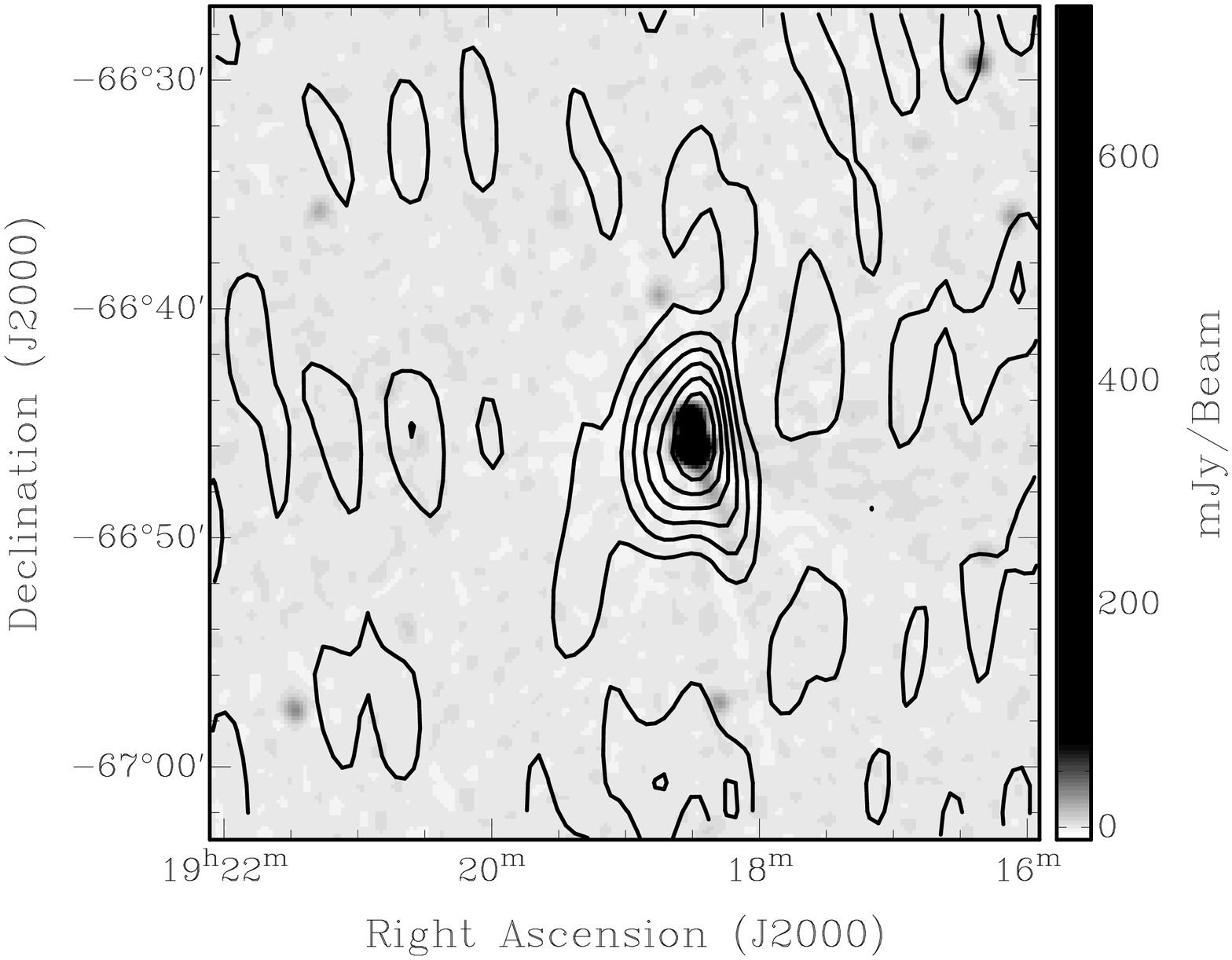,angle=-90,width=0.41\linewidth}}
\caption{ \small MRT contours overlaid on SUMSS image. (a) MRT 
contours before correction and, (b) MRT contours after 
homography-based correction. \normalsize}
\vspace{-10pt}
\label{f:contourplots}
\end{figure}

Figures~\ref{f:contourplots}(a)~and~\ref{f:contourplots}(b) 
show MRT contours before and after correction, respectively, overlaid 
on SUMSS (Sydney University Molonglo Sky Survey) image \citep{mnras:mauch03}, 
for a source around -67$^\circ$ DEC. The corrected MRT image contours in 
Figure~\ref{f:contourplots}(b) overlap well with the source 
in SUMSS image.  

\vspace{-10pt}
\section{Conclusions}
\label{s:results}

The homography-based correction was able to correct for systematics 
in positional errors in the image domain and the errors are now well 
within 10\% of the beamwidth for sources above 15-$\sigma$. The corrected
MRT images are available for download at http://www.rri.res.in/surveys/MRT.

Positional error analysis clearly shows that MRT images are stretched in 
declination (about 1 part in 1000). This translates to a compression of 
the baseline scale, in the visibility domain, which cued towards possible 
errors in our estimation of the array geometry. By formulating a simple 
linear system, using instrumental phases estimated from three well 
separated calibrators whose positions are well known, the array geometry 
was re-estimated \citep{book:thomson}. The estimates show an error of 
about 1~mm/m. In other words, the error is about half a wavelength
at 150~MHz (1~m) for a 1~km baseline. This matches with the observed 
stretching of MRT images.

\vspace{-10pt}
\section*{\small Acknowledgment}
\small S. Daiboo acknowledges a PhD bursary from the South African 
Square Kilometer Array project.

\end{document}